# Dynamics of Na Ion in the Amorphous $Na_2Si_2O_5$ Using Quasielastic Neutron Scattering and Molecular Dynamics Simulations

Mayanak K. Gupta[1,$], Sanjay K. Mishra[1], Ranjan Mittal[1,2,*], Baltej Singh[1,2], Prabhatasree Goel[1], Sanghamitra Mukhopadhyay[4], Rakesh Shukla[3], Srungarpu N. Achary[2,3], Avesh K. Tyagi[2,3], and Samrath L. Chaplot[1,2]

[1]*Solid State Physics Division, Bhabha Atomic Research Centre, Mumbai, 400085, India*
[2]*Homi Bhabha National Institute, Anushaktinagar, Mumbai 400094, India*
[3]*Chemistry Division, Bhabha Atomic Research Centre, Mumbai, 400085, India*
[4]*ISIS Neutron and Muon Facility, Rutherford Appleton Laboratory, Chilton, Didcot, Oxon OX11 0QX, UK*

Corresponding Authors Email: mayankg@barc.gov.in[$], rmittal@barc.gov.in[*]

We have investigated the dynamics of Na ions in amorphous $Na_2Si_2O_5$, a potential solid electrolyte material for Na-battery. We have employed quasielastic neutron scattering (QENS) technique in the amorphous $Na_2Si_2O_5$ from 300 to 748 K to understand the diffusion pathways and relaxation timescales of Na atom dynamics. The microscopic analysis of the QENS data has been performed using ab-initio and classical molecular dynamics simulations (MD) to understand the Na-ion diffusion in the amorphous phase. Our experimental studies show that the traditional model, such as the Hall and Ross (H-R) model, fairly well describe the diffusion in the amorphous phase giving a mean jump length of ~3 Å and residence time about 9.1 picoseconds. Our MD simulations have indicated that the diffusion of $Na^+$ ions occurs in the amorphous phase of $Na_2Si_2O_5$ while that is not observed in the crystalline orthorhombic phase even up to 1100 K. The MD simulations have revealed that in the amorphous phase, due to different orientations of silicon polyhedral units, accessible pathways are opened up for $Na^+$ diffusions. These pathways are not available in the crystalline phase of $Na_2Si_2O_5$ due to rigid spatial arrangement of silicon polyhedral units.

# I. INTRODUCTION

Clean and green storage of energy is achievable by various energy storage technologies[1-4]. Batteries are one of the most eligible devices[5-8] for this purpose. The solid state batteries[6, 9-13] offer a most compact and safe version to store energy at various scales. These batteries contain a solid electrolyte between two electrode materials and hence no separator is required to fabricate such batteries. The solid electrolyte[14], being more stable[6, 9-11] than liquids, could resolve the issues of leakage, chemical stability, vaporization, flammability and dendrites formation to a great extent. Also, they enable the use of metallic Li or Na as anode material[9, 15] and enhanced current charge capacities.

A common feature to the most solid electrolyte materials is availability of sufficient number of lattice sites for the diffusing ions in the solid[16, 17]. Diffusion is made possible by the correlated forward backward hoping processes which give rise to the so called "universal dielectric response". There exist various diffusion mechanisms depending on the structural framework and nature of diffusing ions[18]. The ionic transitional motion is enhanced[19, 20] by rotational motion of transitionally static counter ions as found in the case of solid ionic conductor $Li_2SO_4$[17].

In search of an efficient and cost effective Na and Li ion conductors for solid state Na or Li-ion batteries there have been significant experimental and theoretical studies available in literature[21-28]. The diffusion of Li ions in these have been addressed by several experimental and theoretical modes, like tracer migration, positional displacement as well as electrical property measurements. The positional displacement and site disorder of mobile ions often causes liquid like sublattice in the frame of the rigid lattice. The crystalline $LiAlSiO_4$ (β-eucryptite) and $Li_2X$ (X=O, S, Se) are known to exhibit Li ion conductivity in the temperature range from 800 to 1400 K[25, 29, 30]. The inelastic neutron scattering spectra of β-eucryptite at high temperature show significant broadening in the spectral regimes which is contributed by Li atoms[30]. The Li diffusion in β-eucryptite arises from highly correlated 1-D motion of Li channels along hexagonal c-axis[29]. The glass-ceramics and amorphous solids like $Li_{3+x}PO_{4-x}N_x$(LiPON), $A_7P_3S_{11}$, $A_3OCl$ and $Li_2S–P_2S_5$ have recently attracted attention[23, 31-35] due to their much higher ionic conductivities at room temperature than their equivalent crystalline forms. The higher ionic conduction in amorphous phase than that in the crystalline phase might be due to creation of significant number of percolative pathways with minimal energy barriers for ion migration, which is essentially a geometrical effect due to amorphization that promotes the ionic conduction.

Following the discovery of superionic Na-ion conduction in layered β-alumina, many Na-ion conductors have been discovered[37-40]. $Na_2Si_2O_5$ is known to show Na ion conduction in its amorphous phase with ionic conductivity of 0.01 S/cm$^2$ at 500 °C [41]. At ambient pressure, $Na_2Si_2O_5$ crystallizes in four modifications. The β (monoclinic phase[42] with space group P2$_1$/a), γ (body centered tetragonal phase[43], I4$_1$/a) and δ (monoclinic phase [44] with space group *P2$_1$/n*) phases are metastable, however the only stable phase is known as α-$Na_2Si_2O_5$ (orthorhombic phase[45], Pcnb). X-ray diffraction measurements, differential scanning calorimetry analysis and electrochemical impedance spectroscopy show that the amorphous $Na_2Si_2O_5$ is metastable and transforms to a crystalline[46] form at around 773 K. The crystalline $Na_2Si_2O_5$ is observed to be poor ionic conductor[41, 46]. Moreover, the amorphous phase is structurally and electrically more stable in reducing atmosphere than in oxidizing ones. The presence of H or Al impurities in amorphous $Na_2Si_2O_5$ increases the crystallization temperature[46]. The ionic size of Na in $Na_2Si_2O_5$ is larger as compared to that of Li in $Li_2Si_2O_5$. However, the molecular dynamical simulations report[41, 47] a lower activation energy in $Na_2Si_2O_5$ (0.3 eV) than that in $Li_2Si_2O_5$ (0.47 eV). The *ab-initio* simulations[41] on amorphous $Na_2Si_2O_5$ show that Na preferentially diffuses within the layered channels formed by corner-shared $SiO_4$ tetrahedra.

The quasielastic neutron scattering (QENS) and molecular dynamics simulations are extensively used for studying the diffusion phenomenon at various length and time scale, for example diffusion of organic/inorganic solvents (or polymers) in various zeolites frameworks[48, 49], dynamics of metal organic framework (MOF) during adsorption of gases[50], and dynamics in molten salts[51]. The QENS spectrum are measured as a function of temperature and momentum transfer. The spectroscopic technique enables to study the dynamics (segmental relaxation) of diffusing atoms or group of atoms following non-Debye and non-Arrhenius temperature dependence[49].

Further, QENS technique is also very well used to measure the dynamical motion of ions in materials[52]. In case of solid electrolyte, the technique was first time used to measure the ionic diffusion in β–alumina[53]. These experiments[53] revealed two types of diffusion mechanisms of Na at 400ºC viz. localized jumps and long range diffusion. The QENS study of α-phase of $Mg_3Bi_2$ provides the mechanism of superionic diffusion of $Mg^{+2}$ in the rigid bismuth lattice[54]. The diffusion coefficient is estimated as 2.7±0.3 × 10$^{-9}$ m$^2$/s. The modelling of *Q* dependence of the half-width at half-maxima of the quasielastic broadening implies a cross over from jump like diffusion at small distance to continuous diffusion at longer distances.

Recently high neutron flux of OSIRIS time of flight spectrometer at ISIS,UK has been used to measure Na-ion diffusion in sodium cobaltate[55].We have performed QENS measurements over a range of temperature from 300-773 K to evaluate the diffusion constant, nature and pathways of Na-ion diffusion in the amorphous phase of $Na_2Si_2O_5$. The QENS measurements show a significant broadening of the elastic line beyond the resolution of the instrument. The molecular dynamics simulation using both the classical and ab-initio density functional theory methods have also been performed for microscopic understanding and corroborating the experimental observations.

## II. EXPERIMENTS

The QENS experiments are performed from amorphous phase of $Na_2Si_2O_5$. First α-$Na_2Si_2O_5$ was synthesized[45] by conventional solid state reaction between $Na_2CO_3$ and $SiO_2$ with four step heating protocol. Preheated $Na_2CO_3$ and $SiO_2$ in 1.05:1.0 molar ratio was thoroughly ground in acetone media and dried in air. The dried powder mixture was heated at 550°C for 12 hours and then reground. This homogenous mixture was pelletized and heated at 650 °C, 750 °C and 800 °C in air, for 12h at each temperature. The crystalline sample of α-$Na_2Si_2O_5$ was obtained after heating 800°C for 12h. The amorphous $Na_2Si_2O_5$ was prepared by melt quench method. Crystalline sample of α-$Na_2Si_2O_5$ was heated at 950 °C in a platinum crucible and held for about 4h and then quenched by dipping crucible in liquid nitrogen. The obtained transparent glass sample was finely ground to powder and characterized by X-ray diffraction.

The QENS measurements were performed on polycrystalline sample of $Na_2SiO_5$ (amorphous) using the OSIRIS[56-58], the indirect geometry time of flight spectrometer at ISIS. The final energy in the measurements was fixed at 1.84 meV using pyrolytic graphite (002) analysers, giving the energy resolution (HWHM) of 12.7μeV. Scattered neutrons were detected over an angular range of 2θ = 25–160°, providing available Q range 0.42–1.85 Å$^{-1}$. The amorphous polycrystalline powder samples were kept in a niobium sample holder and sealed. A high temperature furnace operating in high vacuum served as the sample environment. The experimental data on sample has been recorded at several temperatures from 300 to 773 K. The data taken on empty niobium sample holder are used to remove the background scattering due to the sample environment. A vanadium standard was measured for normalization of detectors and for measuring the resolution of the instrument. The data as collected in various detectors were grouped in the wave-vector (Q) steps of 0.05 Å$^{-1}$ for improving the counting statistics. The experimental data were normalized and corrected for detector efficiency and detailed balance using the standard program package for data analysis available at the spectrometer. All data analyses were performed using the QENS data analysis interface as implemented in Mantid software[59, 60].

## III. CALCULATIONS

The molecular-dynamics simulations of the diffusion process are carried out using both the ab-initio density-functional theory and classical approach based on inter-atomic potentials. These two approaches were necessary since the DFT simulations can only be performed, even on supercomputers available to us, on a rather small simulation cell that may limit the pathways of the diffusion process, while the classical simulations could be performed on much larger cells and also for much longer time duration. We have also performed the lattice dynamics calculations using both the approaches to evaluate the phonon spectrum. As the results from the two approaches are found to be consistent with each other, we have used the larger simulations to visualize the geometries of the diffusion pathways and their time scales.

For the classical simulations, we have used the Buckingham model of inter-atomic potential, which consists of long-range Coulomb interaction, short-range repulsive interaction and van der Waals interaction terms, given by

$$V(r) = \frac{e^2}{4\pi\varepsilon_o} \frac{Z(k)Z(k')}{r} + a \exp\left[\frac{-br}{R(k)+R(k')}\right] - \frac{C}{r^6} \qquad (1)$$

where, r is the separation between the atoms of type k and k', and R(k) and Z(k) are, respectively, the effective radius and charge of the $k^{th}$ atom. As in earlier studies, a= 1822 eV and b=12.364 have been treated as constants. This choice has been successfully used earlier to study the phonon properties of several complex solids[61, 62]. The optimized parameters as used in our calculations are Z(Na)=1.0, Z(Si)= 2.6, Z(O)= -1.44, R(Na)=1.07 Å, R(Si)= 0.70 Å, R(O)= 2.09 Å. The van der Walls interaction (C = 100.0 eVÅ$^6$) is introduced only between the oxygen atoms. The phonon calculations in the α-$Na_2Si_2O_5$ are performed using the software[63] developed at Trombay.

The classical molecular dynamics simulations have been performed on a super cell size of 4×5×2 (1440 atoms) for 200 pico-seconds after the equilibration. The simulations are carried out in NPT ensemble. To create the amorphous phase, we have melted the crystalline phase at 2500 K in 10 pico-second and then quenched to 10 K. The quenched structure is used as amorphous phase of $Na_2Si_2O_5$. For simulation in amorphous phase, we have equilibrated the system for 300 pico second to assure that there should not be any energy or temperature drift due to strain in amorphous structure. To characterize the amorphous phase, we have performed simulation at 300K in crystalline and amorphous phase for 200 pico-seconds after equilibration. The simulations in both the crystalline and amorphous phases have been performed at various temperatures from 300K to 1200 K using DL_POLY software[64].

To perform ab-initio density functional theory based phonon calculation we have used Vienna based ab-initio simulation package[65, 66] (VASP). The supercell scheme has been used for phonon calculation, i.e. the forces on various atoms have been calculated on a 2×2×1 supercell of dimension. The supercell and subsequent atomic displacements have been generated using PHONOPY software[67]. Further, the forces and displacement of different configuration have been used to calculate the phonon spectrum using PHONOPY software. The total energy and forces calculation using VASP are performed using the projected augmented wave (PAW) formalism of the Kohn-Sham density functional theory within generalized gradient approximation (GGA) for exchange correlation following the parameterization by Perdew, Becke and Ernzerhof[68, 69]. The plane-wave pseudo-potential with a plane-wave kinetic energy cutoff of 700 eV was adopted. The integration over the Brillouin zone is sampled using a k-point grid of 6×6×2, generated automatically using the Monkhorst-Pack method[70]. The total energy is minimized with respect to structural parameters. The convergence criteria for the total energy and ionic forces were set to $10^{-8}$ eV and $10^{-4}$ eVÅ$^{-1}$, respectively.

For ab-initio molecular dynamics (MD) simulation, we have taken a single k-point in the Brillouin zone. An energy convergence of $10^{-6}$ eV has been chosen for self-consistence convergence. A time step of 2 femtosecond is used. The amorphous structure in the ab-initio method has been generated using the following steps. We started with the crystalline $Na_2Si_2O_5$ phase (216 atoms, 2×3×1 supercell of the orthorhombic phase) and melted it using ab-initio MD simulation at 2500 K for 10 picoseconds. Then we quenched the structure to 10 K and used the quenched structure as an amorphous structure of $Na_2Si_2O_5$. Simulations are performed for a series of temperatures from 300 to 1100 K. Initially, the structure was equilibrated for 10 ps to attain the required temperature in NVT simulations through a Nose thermostat[71]. Then for the production runs up to 40 ps, NVE simulations are performed. At each temperature, a well-equilibrated configuration is observed during the 40 ps simulation.

The isotropic diffusion coefficient can be estimated from the time dependence of mean square displacement as given below:

$$D = \langle u^2 \rangle / (6\tau) \quad (2)$$

where $\langle u^2 \rangle$ is the change in the mean square displacement in time $\tau$. The MSD at time $\tau$ is calculated using the following equation[22, 72]

$$u^2(\tau) = \frac{1}{N_{ion}(Nstep-N\tau)} \sum_{i=1}^{N_{ion}} \sum_{j=1}^{N_{step}-N_\tau} |r_i(t_j + \tau) - r_i(t_j)|^2 \quad (3)$$

Here $r_i(t_j)$ is the position of $i^{th}$ atom at $j^{th}$ time step. $N_{step}$ is total number of simulation steps and $N_{ion}$ is total number of atoms of a particular type in the simulation cell. $N_\tau = \tau/(\delta t)$, where $\delta t$ is step size.

## IV. RESULTS AND DISCUSSION

### A. Quasielastic Neutron Scattering

QENS is a powerful experimental technique, which enables to measure diffusion process. It exhibits the broadening of the elastic scattering caused by non-periodic "diffusive motion" in the sample as a function of momentum transfer. The quasielastic signal in neutron scattering experiments contains the sum of coherent and incoherent contributions. In case of $Na_2Si_2O_5$, silicon and oxygen only scatter coherently, the incoherent contribution is dominant by the scattering from sodium. Sodium has nearly identical coherent and incoherent cross sections[73]. The dynamical motion of Na-diffusion using QENS experiments in this potential battery material is challenging due to low neutron scattering cross-section of Na[73], and the data interpretation is also difficult due to the coherent nature of the scattering from Na atom.

**Fig 1** shows comparison of the dynamic neutron scattering function $S(Q,\omega)$ of amorphous $Na_2Si_2O_5$ integrated over all Q and at 300 K and 748 K and instrument's resolution. The measurements at 748 K provide clear evidence of the QENS broadening. The data at 748 K can be fitted to one Lorentzian peak and one delta function convoluted with the resolution function of the instrument providing evidence of QENS broadening from the sample. The diffraction data (**Fig 2**) on the same temperature at 300 to 748

K shows that in vacuum, we did not found any evidence for crystallization of sample and provide evidence that the sample remains in the amorphous phase in the entire temperature range of our measurements. The temperature dependent variation of elastic -intensity extracted from dynamic neutron scattering function summed over all Q is shown in **Fig 3**, which does not show any sudden drop of elastic intensity. This is consistent with the fact that there is no phase change in the solid at the experimental temperature range. However, from Fig. 3, the change in elastic intensity is more than that expected from only the Debye-Waller factor, which could be due to increase in disorder[74].

Further in order to understand the microscopic nature of sodium diffusion we need Q dependence of quasielastic broadening. The data collected in various detectors in the Q range from 0.6 to 1.7 Å$^{-1}$ were grouped in the Q steps of 0.05 Å$^{-1}$. The individual data in steps of 0.05 Å$^{-1}$ are fitted to one Lorentzian peak and one delta function convoluted with the resolution function of the instrument. All the fits were done over an energy range -0.4≤E≤0.4 meV, chosen to be symmetric around E=0. The Q-dependent variation of elastic peak height, amplitude and width of Lorentzian peak extracted from dynamic neutron scattering function is shown in **Fig. 4**. It can be seen (**Fig. 4)** that peak height of elastic and amplitude of Lorentzian peak increase with increasing Q. The monotonic increase in intensity of the elastic contribution with Q over 0.6 to 1.7 Å$^{-1}$ might be partially due to the amorphous structure and partially due to coherent scattering. However, width of Lorentzian peak which gives information about the diffusion (quasielastic line width) shows a significant Q dependence and indicates increment of the diffusion process.

The Q dependence of HWHM is fitted using the Hall and Ross (H-R) model[75, 76] as well as the Chudley-Elliott (C-E) model[77], which are appropriate models for variable length jump diffusion process. In case of H-R model[75, 76], the Q dependence of HWHM ($\Gamma(Q)$) is related to the mean jump length (d) and average jump time ($\tau$) from one site to another site by following expression:

$$\Gamma(Q)=(1-\exp(-<d^2>Q^2/6))/\tau \qquad (4)$$

Similarly, for Chudley-Elliott model[77] the corresponding expression is given by:

$$\Gamma(Q)=(1-\mathrm{Sin}(Qd)/Qd)/\tau \qquad (5)$$

**Fig 4** shows the Q dependence of the QENS broadening with a fit of the H-R and Chudley-Elliott models of jump diffusion for amorphous $Na_2Si_2O_5$ at 748 K. The fit of the H-R model to QNES data gives a mean residence time 9.1 ps (13.8 meV$^{-1}$) and diffusion jump-length of 3.0±0.8 Å, which give a diffusion coefficient of 16.5 ×10$^{-10}$ m$^2$/sec. Similarly the Chudley-Elliott model gives τ and d values of 11.4 ps (17.4 meV$^{-1}$) and 3.2 ±0.5 Å respectively, resulting in diffusion coefficient value of 15.0 × 10$^{-10}$ m$^2$/sec. As expected, the jump-length as estimated from the fits of the experimental data matches very well with the calculated first-neighbor distance of 2.98 Å in the amorphous phase (discussed below).

## B. Molecular Dynamics Simulations

The magnitude of broadening in the QENS peak depends on the diffusion coefficient of the diffusive species; in our case it is Na-ion. The quasi-elastic broadening as observed from experiment gives quantitative estimate of the diffusion coefficient. However, the nature of diffusion and microscopic view may not be completely inferred from the experiment. Hence in order to get the insight of Na- ion diffusion, we have performed simulation in the amorphous phase of $Na_2Si_2O_5$ compound. In order to simulate the quasi-elastic broadening with appropriate momentum resolution, the simulation is performed for few hundred pico-seconds on large supercell (30 Å cell dimension gives ~ 0.2 Å$^{-1}$ momentum transfer resolution). Such a big scale simulation for a set of temperatures is extremely expensive with ab-initio based molecular dynamics method and but easily manageable with classical molecular dynamics methods. The main challenge of classical methods is a good set of potential parameters, which can be used for calculation of thermodynamical and transport properties. Further, we have also performed the ab-initio molecular dynamics simulation in the amorphous phase of $Na_2Si_2O_5$.

The amorphous structure of $Na_2Si_2O_5$ consists of randomly oriented $SiO_4$ tetrahedral units and stuffed with Na. In order to optimize the potential parameter of the Buckingham potential, we have performed rigorous calculations with various set of parameters. We have tested several sets of parameters and chosen the set of parameters which reproduces the closest experimental structure and phonon dynamics as obtained from DFT calculations (**Fig 5**) in the α-$Na_2Si_2O_5$. The ab-initio calculated phonon density of states shows phonon frequencies range extended up to 140 meV. The potential model approach also shows that phonon spectrum is up to 140 meV. The phonon spectrum as calculated from interatomic potential, up to about 95 meV, is qualitatively in fair agreement with ab-initio phonon density of states. However, we find that the in the ab-initio calculated phonon spectrum, there is a phonon band gap around 95-115 meV, while in potential model approach there is no such phonon band gap. This might be due to the fact that oxygen shell charge/polarizability is not taken care in our potential model. In order to include all these effects, the potential model required more parameters and would not be suitable for very large simulation. It should be noted that thermodynamic behavior of compounds is largely dominated by low energy spectrum of phonon. Hence the optimized set of potential parameters can be further used in the molecular dynamics simulation to investigate the dynamics of atoms at high temperatures. The close reproducibility of structure and phonon dynamics satisfies the static and dynamics equilibrium condition.

The computed time averaged pair-distribution functions of various pairs of atoms in $Na_2Si_2O_5$ are shown in **Fig 6** in the amorphous and crystalline phases at 300 K. It can be seen that total pair distribution function (PDF) among atoms using the two approaches of ab-initio and potential model show fair agreement with each other. We found that the crystalline structure consists of sharp and well defined peaks at various bond lengths representing its long range order, while in the amorphous phase the sharp peak is seen only for the short range order of the first neighbors, as expected in amorphous solids[78]. It shows that Si-O PDF intensity for the first neighbor distance ~1.6 Å does not change much between the crystalline and amorphous phase, while other higher neighboring peaks shows significant change. This is due to the fact that the polyhedral unit of $SiO_4$ does not change its shape, while its orientation (medium range order) changes significantly in the amorphous phase. Further the Na-Na PDF intensity changes significantly in the amorphous structure with respect to the crystalline structure even for the short range order, which signifies that the Na distribution changes significantly between the crystalline and amorphous phases. The broad structures at higher distance are attributed to the broad distributions of inter-atomic distances in medium and long range, which indicates a well-constructed amorphous structure. We have performed further analysis with this configuration.

The classical as well as ab-initio molecular dynamics simulations have been performed at various temperatures from 300 K to 1100 K. Further, we have analyzed the mean squared displacement (MSD) of various atoms in the crystalline and amorphous phase. It is interesting to note that the diffusion in $Na_2Si_2O_5$ occurs only in the amorphous phase while crystalline phase does not show any diffusion up to 1100 K. We found that the mean square displacement (MSD) of Na atoms shows significant time dependence at and above 800 K in the amorphous phase while in the crystalline phase even up to 1100 K there is insignificant variation of MSD with time. This gives a signature that in the amorphous phase significant diffusion occurs at and above 800 K while there is no diffusion in the crystalline phase.

In **Fig 7**, we have shown the MSD obtained from ab-initio and classical molecular dynamics simulation in the amorphous phase at various temperatures. The mean squared displacements of various atoms have been calculated using equation (3). We found that the mean square amplitude of Na atoms keeps increasing with time, which is a signature of the diffusion of Na atoms in the compound (**Fig 7**). The mean squared displacement of other atoms, Si and O, does not show any increase with time, hence there is no significant diffusion of these atoms, which is essential for a stable solid electrolyte. The diffusion coefficient as estimated from ab-initio calculations (on a 216 atoms cell) is smaller in comparison to that from the classical calculations (on a 1440 atoms cell) (**Fig 8**). This might be due to the finite size effect in molecular dynamics simulation. As discussed below, we found that the ab-initio simulations are useful to visualize the geometries of the diffusion pathways and their time scales, which is one of the important considerations of this work.

The time dependence of mean squared displacement of various atoms has been used to estimate the diffusion coefficient (**Fig 7**). The calculated diffusion coefficients at various temperatures as estimated from ab-initio and classical molecular dynamics simulation are shown in **Fig 8**. The energy barrier for sodium diffusion can be estimated from the fitting of the temperature dependence of diffusion coefficients with Arrhenius relation i.e.

$$D(T) = D_0 \exp(-E_a/K_B T) \qquad (6)$$

One can linearize this equation by taking log of it, i.e.,

$$\ln(D(T)) = \ln(D_0) - E_a/K_B T \qquad (7)$$

Here $D_0$ is the constant factor, $K_B$ is the Boltzmann constant and T is temperature. We have fitted the simulation data to equation (5) in Fig. 8 for the classical and ab-initio MD simulations and the value of the activation energy is found to be 0.51 eV and 0.41 eV from classical and ab-initio MD simulations respectively.

The estimated diffusion coefficients (**Fig 8**) show very small diffusion up to 800 K, and above that the diffusion coefficient shows a huge jump. To understand how the Na ion diffusion coefficient increases with temperature, we have plotted the mean squared displacement of individual Na atoms (**Fig 9**). We observe that a few Na atoms show sudden jump in the mean squared displacement, which essentially indicates a jump like diffusion of Na atoms in the amorphous phase. At 700 K, only a few Na atoms show jump in MSD with a jump length of ~2 Angstrom, however, at above 900 K, significant number of Na atoms show jump in MSD. Further, the jump length shows a distribution from ~1.4 Å to 3 Å. This distribution of the jump-lengths is a consequence of the amorphous structure of materials where we have an irregular spatial distribution of Na atoms. In **Fig. 10,** we have shown the distribution of Na-Na distance in the amorphous structure with different magnitude of bond length denoted by a different color. We can see how the Na atomic sites are connected in the amorphous phase and provide the pathways for diffusion. Further, it may be noticed that even the same path length may not result in the same activation energy, since for the same path length the ionic trajectory for Na hopping may be different which depends on the local site environment.

The calculated trajectories of various atoms from the ab-initio MD simulation in the amorphous phase are shown in **Fig 11**. We find that the Na atoms move from one site to another via jump like diffusion, and the number of Na atoms participating in jump diffusion as well as jump frequency increases significantly with temperature. The increase in the number of Na atoms and jump frequency is gradual, which might be due to fact that in the amorphous phase the distance between two Na sites are not unique and show a distribution.

In order to gain microscopic understanding of the diffusion behavior from quasi-elastic broadening as obtained from neutron data, we have computed incoherent dynamical structure factor[48] $S(Q, \omega)$ at allowed wavevector Q values at 1000 K. The lowest Q value of 0.2 Å$^{-1}$ in the calculations is limited by the size of the super cell. The calculated dynamical incoherent structure factors $S(Q, \omega)$ are fitted with the single Lorentz peak function

$$S(Q,\omega) = A \frac{\Gamma}{\Gamma^2 + \omega^2} \qquad (8)$$

Here A is the normalization constant and $\Gamma$ is half width at half maximum energy (HWHM).

The calculated HWHM is plotted as a function of wavevector Q at 1000 K ($\Gamma(Q,T)$) for amorphous phase of Na$_2$Si$_2$O$_5$ (**Fig 12**). This contains the information about the diffusion behavior in space and time. In amorphous phase the Na atoms jumps from one position to another and it is expected to not be a continuous diffusion as there is no unique jump length in the amorphous structure.

As shown above, both the H-R model and Chudley-Elliott model fits experimental data fairly well. However, our simulated quasielastic broadening with Q fits (**Fig 12**) well with H-R model (eq (4)) of jump diffusion. The fitting using H-R model gives a jump length of about 2.6 Å at 1000 K and residence time of 15.4 ps (23.4 meV$^{-1}$). The jump length matches very well with that obtained from the experimental data of 3.0±0.8 Å and also the first neighbor distance (**Fig. 6**) in the amorphous structure (~2.98 Å), although the residence time is longer than that obtained from the experiments values of 9.1 ps (13.8 meV$^{-1}$). The diffusion coefficient value as estimated using H-R model from calculations at 1000 K is 7.3 × 10$^{-10}$ m$^2$/sec, while the corresponding value from QENS data at 748 K is 16.5 ×10$^{-10}$ m$^2$/sec.

## CONCLUSIONS

Here we have reported the quasielastic neutron scattering measurements of the amorphous Na$_2$Si$_2$O$_5$ as a function of temperature. We have analyzed the data and obtained the quasielastic width as a function of temperature and momentum transfer, which give the time scale and jump length of Na diffusion process in the compound. Our quasielastic neutron scattering measurements show a typical jump length of ~3 Å and residence time is about 9.1 picoseconds. The molecular dynamics simulations are useful for further insight of the diffusion mechanism and the simulations show that there is a distribution of jump length centered around 2.6 Å. The analysis of Na trajectories shows that in the amorphous phase various Na sites have been created due to misorientation of silicon polyhedral units and some of them are connected and provide a pathway for Na diffusion, while these pathways were not available in crystalline phase.

## ACKNOWLEDGEMENTS

The use of ANUPAM super-computing facility at BARC is acknowledged. The authors also thank STFC, UK for the beam-time at ISIS and also availing the travel support from the Newton fund, ISIS, UK.

FIG1 (Color online).(a) Comparison of as observed dynamical neutron scattering function $S(Q,\omega)$ of amorphous $Na_2Si_2O_5$ integrated over all Q at 300 K (Blue line) and 748 K (Red line) and the instruments resolution (vanadium at 300 K; Green Line). The data have been normalized to unity at the peak position. (b) Fit of one Lorentzian peak and one delta function convoluted with the resolution function at a selected Q= 1.2 Å$^{-1}$ slice of $S(Q,\omega)$, and a linear background.

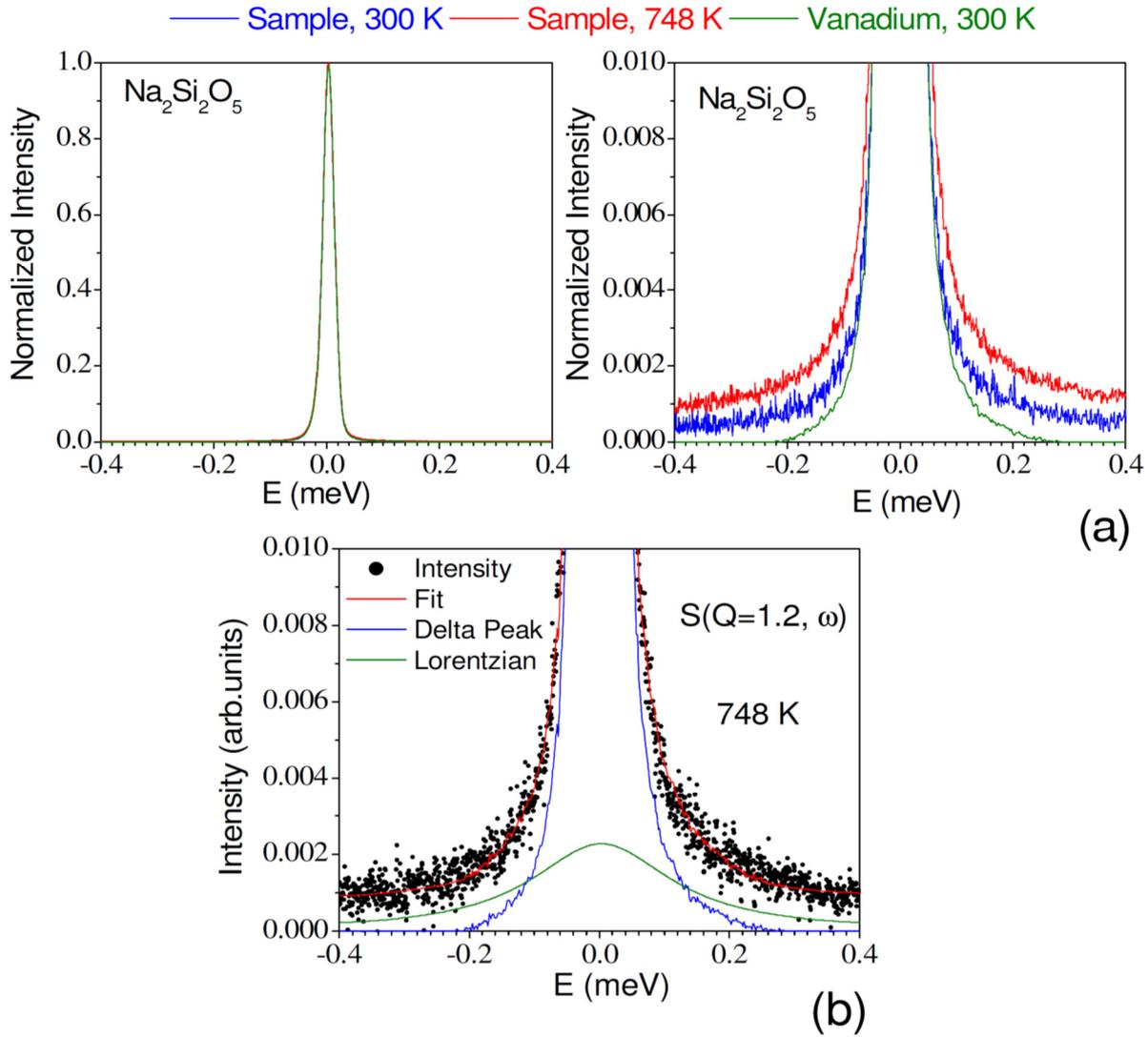

FIG 2 (Color online). The integrated intensity as a function of d-spacing at 300 K and 748 K.

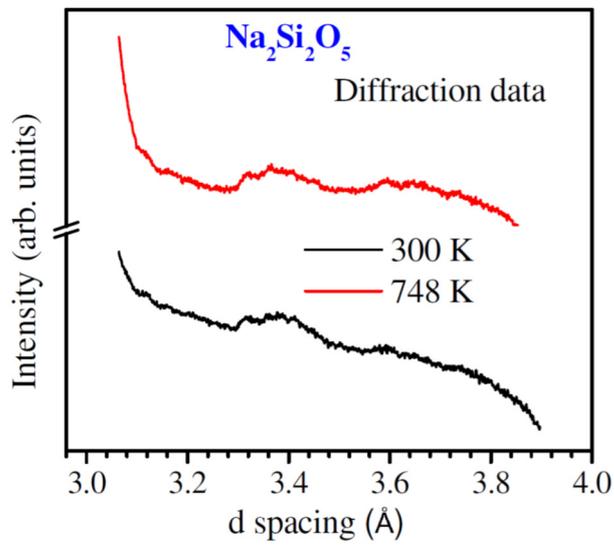

FIG 3 (Color online). Temperature dependent variation of integrated elastic peak intensity extracted from dynamic neutron scattering function $S(Q,\omega)$. The data is summed over all Q.

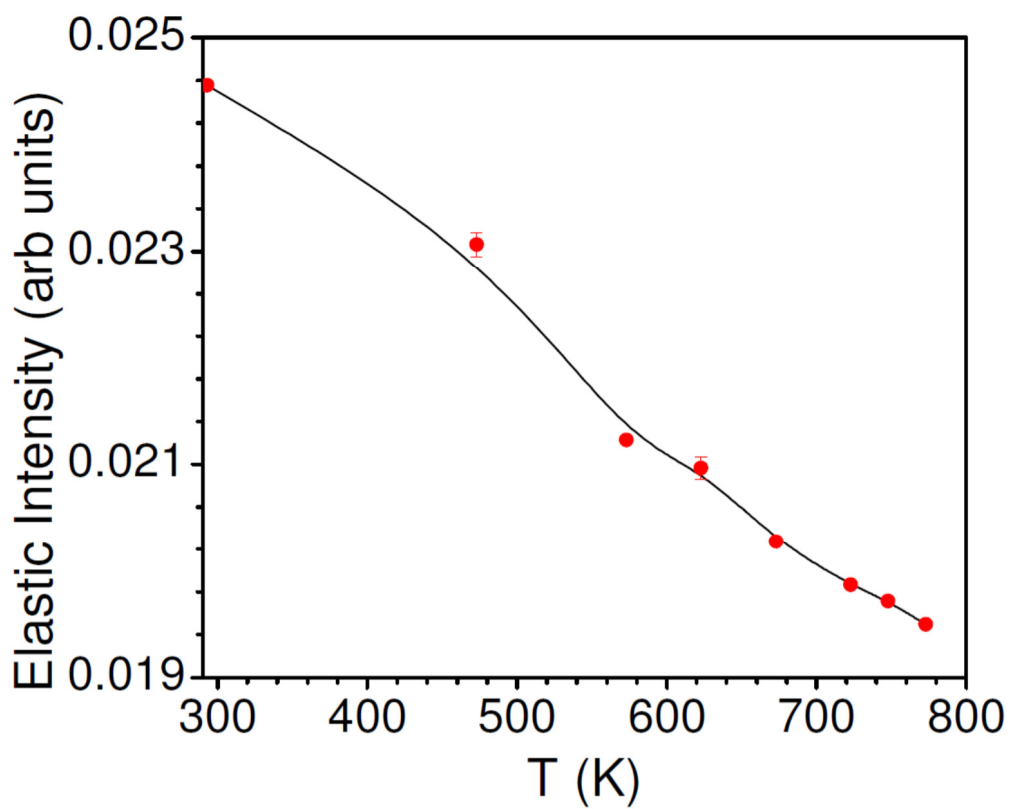

FIG 4 (Color online). Q dependent variation of (a) elastic peak height, (b) amplitude of Lorentzian and (c) and half width at half maximum (HWHM) of Lorentzian peak extracted from dynamic neutron scattering function $S(Q,\omega)$ of amorphous $Na_2Si_2O_5$ at 748 K in vacuum. Q-binning is done in steps of 0.05 Å$^{-1}$. The solid red lines in (c) is obtained from the fits of the Hall and Ross (Red line) and Chudley-Elliott (Blue line) models to the experimental data.

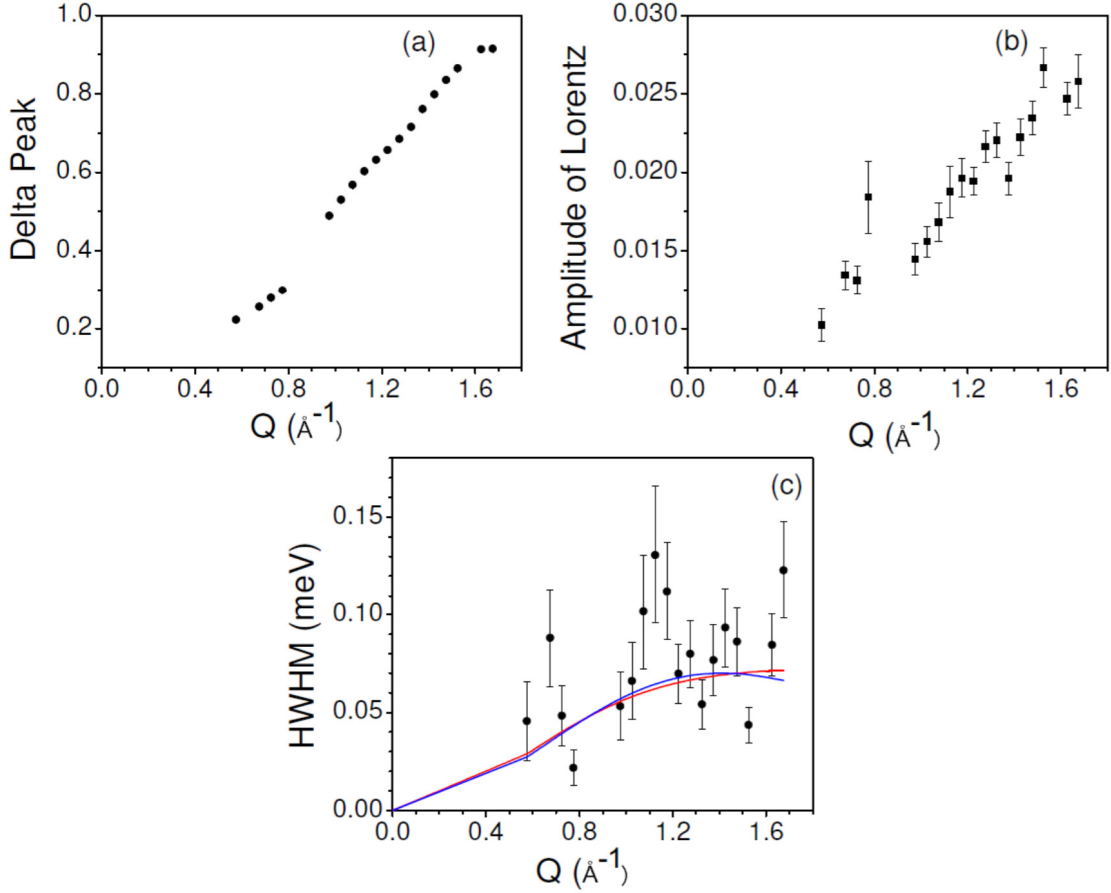

FIG 5 (Color online). The calculated phonon density of states of $Na_2Si_2O_5$ in orthorhombic phase using ab-initio density functional theory method and classical potential approach.

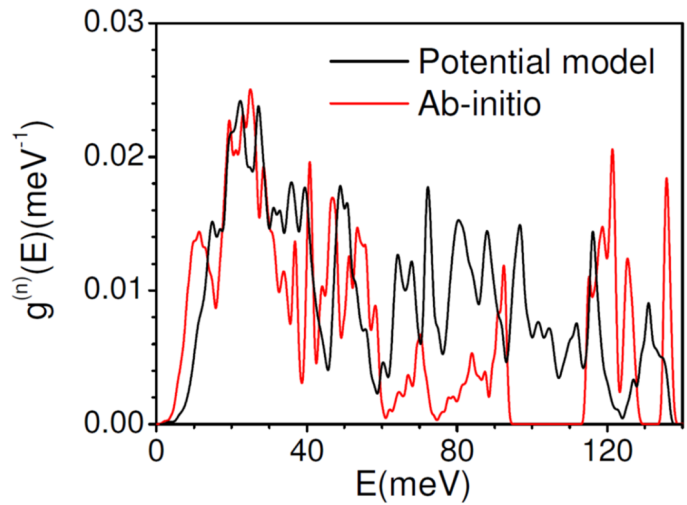

FIG 6 (Color online). The calculated pair distribution functions of various pairs of atoms in crystalline and amorphous phase of $Na_2Si_2O_5$ using ab-initio and classical MD approach.

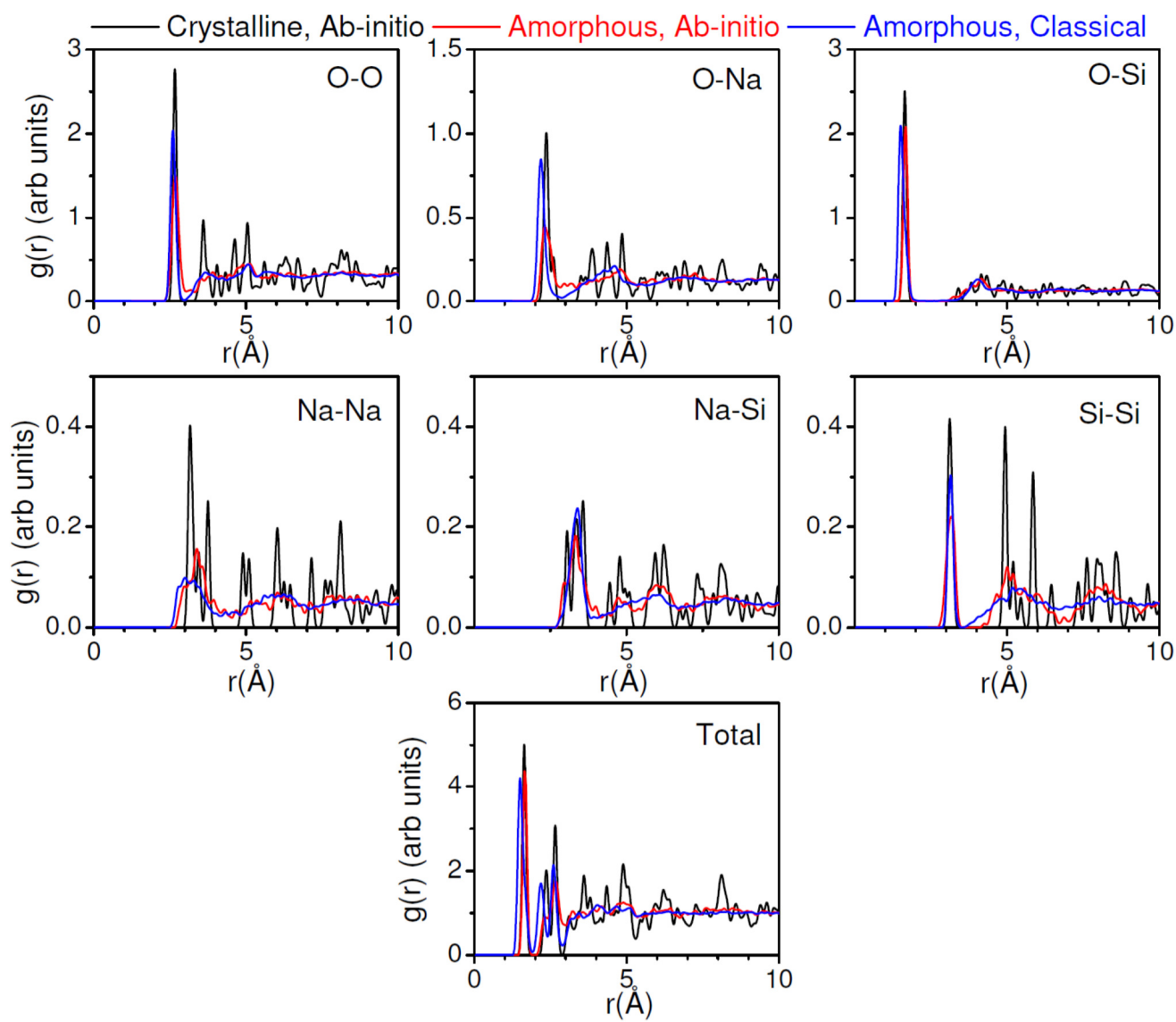

FIG 7 (Color online).The calculated mean squared displacement of various atoms in amorphous phase of $Na_2Si_2O_5$ using classical (Top panel) and ab-initio (Bottom panel) MD simulations.

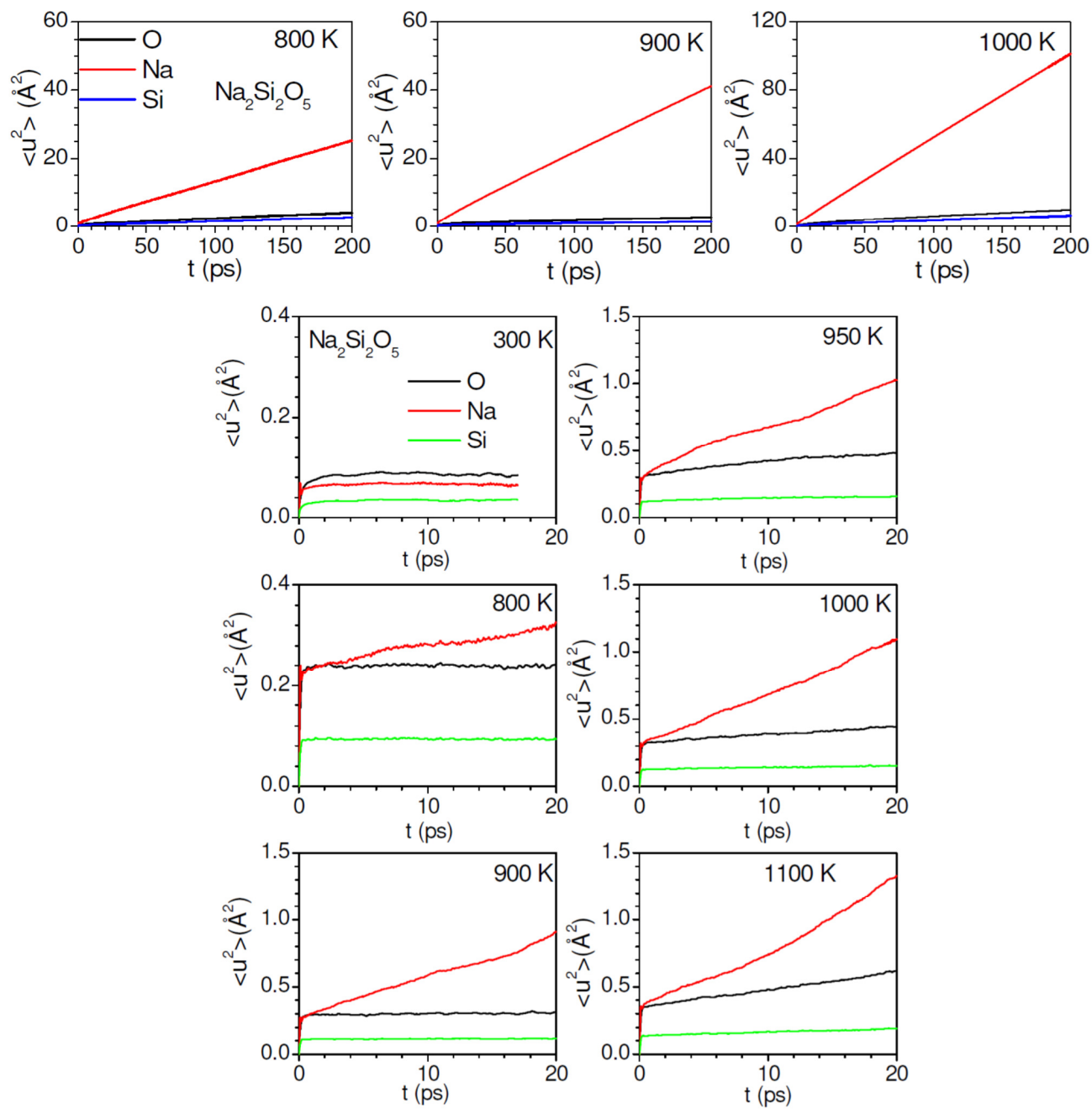

FIG 8 (Color online). The calculated diffusion coefficients and activation energy barriers in amorphous phase of $Na_2Si_2O_5$ as estimated using classical and ab-initio molecular dynamics simulations.

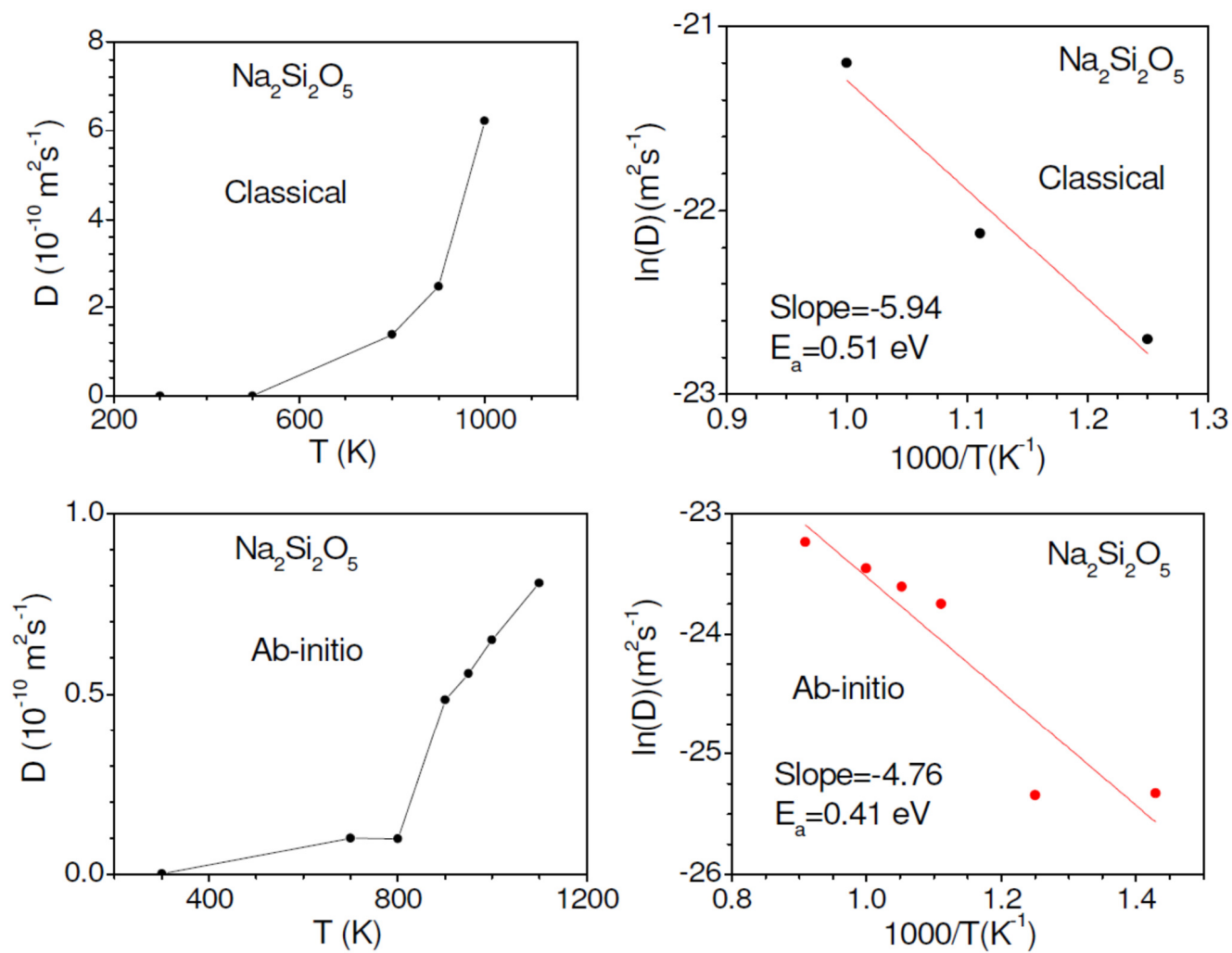

FIG 9 (Color online). The calculated displacement of Na atoms in $Na_2Si_2O_5$ as estimated using ab-initio molecular dynamics simulations.

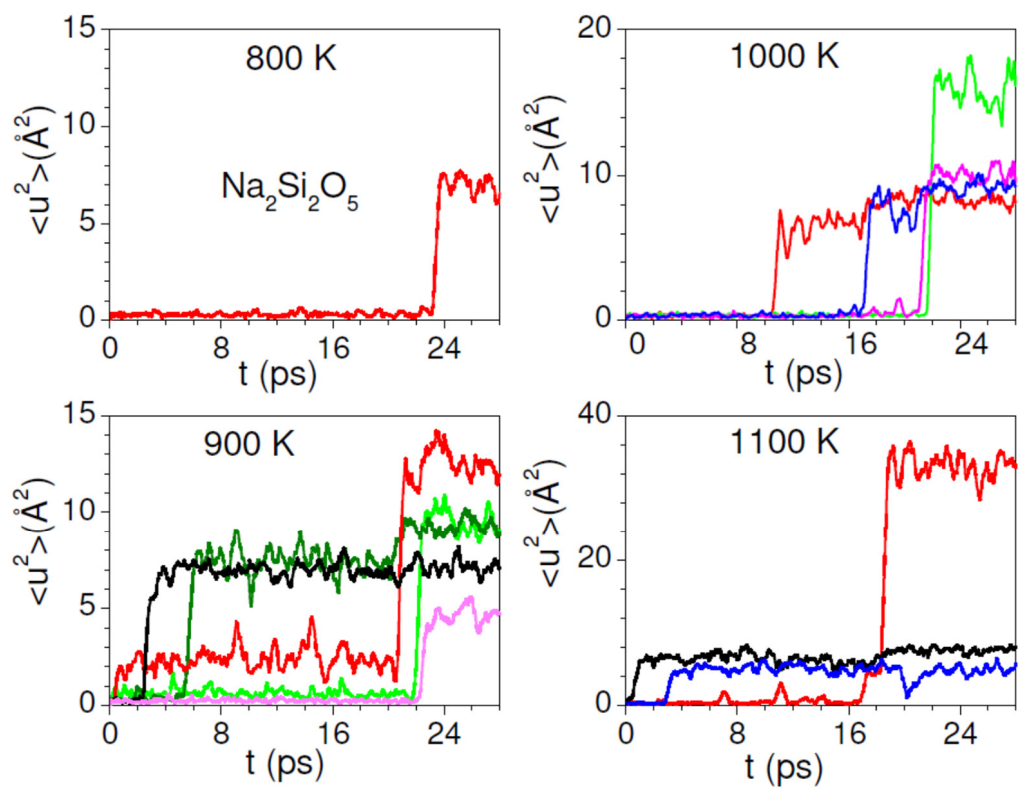

FIG.10 (Color online) The distribution of Na atoms in the amorphous phase, depicted in the ab-initio simulation cell. The color of the bond joining any two Na atoms represents the distance between them (Red= 2.7-3.0 Å; Green 3.0-3.3 Å; Blue 3.3-3.6 Å; Black 3.6-3.9 Å).

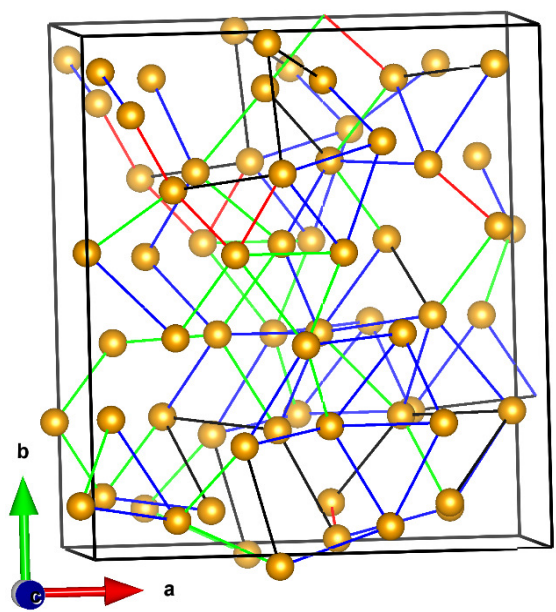

FIG.11 (Color online) The observed trajectories of selected Na atoms in ab-initio MD simulations at 800 K and 900 K in $Na_2Si_2O_5$ in the b-c plane. All the atoms (Key: Na, Yellow sphere; Si, blue sphere; O, red sphere) are also shown at their respective position in the beginning of the simulation. The time dependent positions of the selected Na atoms are shown by green colored dots. The numbers below each frame indicate the temperature of the simulation and atomic fractional co-ordinate of the selected Na atom at the beginning of the ab-initio simulation. The fractional values are given with respect to the simulation cell.

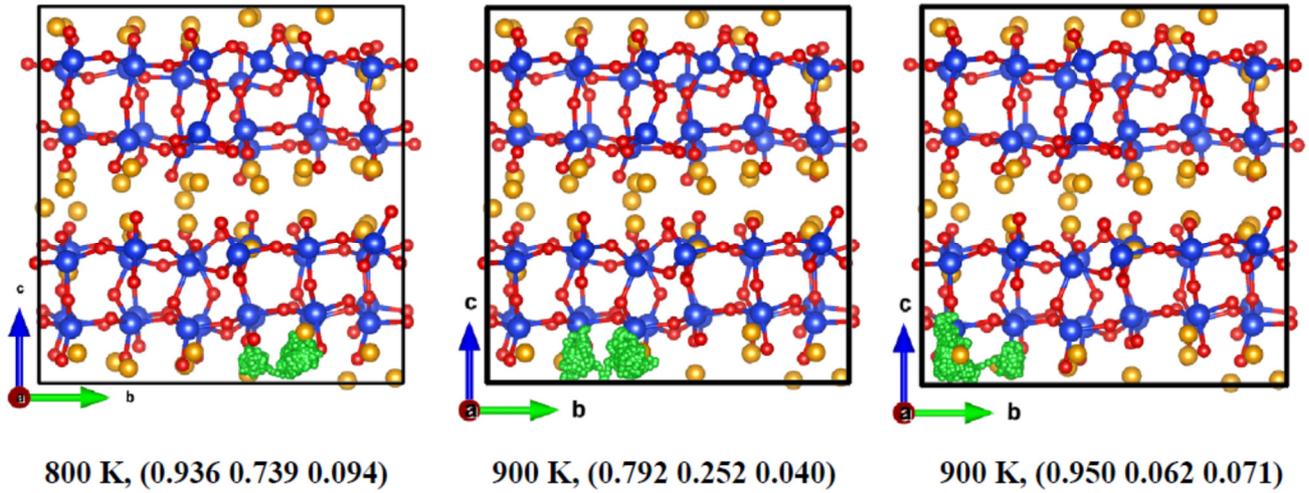

800 K, (0.936 0.739 0.094)   900 K, (0.792 0.252 0.040)   900 K, (0.950 0.062 0.071)

FIG 12 (Color online). The calculated wave vector dependence of half width at half maximum (HWHM) of dynamical incoherent structure factor of Na atoms in amorphous phase of $Na_2Si_2O_5$ using classical MD approach at 1000 K and fitted with Hall and Ross (H-R) and Chudley-Elliott (C-E) models.

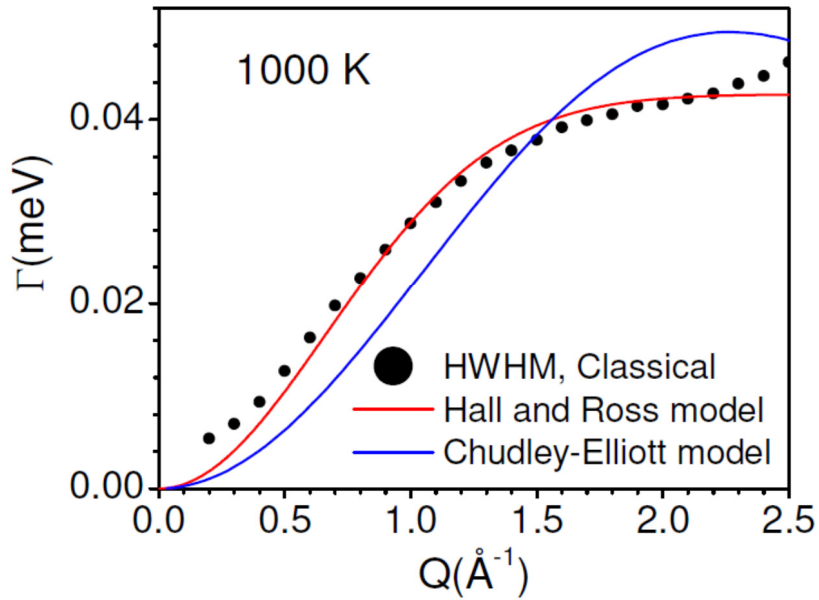